\documentclass{article}
\usepackage[preprint]{icml2026}
\usepackage{amsmath,amssymb,amsfonts}
\usepackage{booktabs}
\usepackage{multirow}
\usepackage{graphicx}
\usepackage{hyperref}
\usepackage{url}
\usepackage{microtype}
\usepackage{array}
\usepackage{caption}
\usepackage[most]{tcolorbox}
\definecolor{darkgray}{gray}{0.3}

\icmltitlerunning{Reflexivity as Prompt}

\begin{document}

\twocolumn[
\icmltitle{Reflexivity as Prompt: Does Awareness of Self-Reinforcing \\Market Dynamics Improve LLMs as Financial Market Forecasters?}

\icmlsetsymbol{equal}{*}
\begin{icmlauthorlist}
\icmlauthor{Eugene Park}{mit}
\end{icmlauthorlist}
\icmlaffiliation{mit}{Massachusetts Institute of Technology}
\icmlcorrespondingauthor{Eugene Park}{ewp@mit.edu}

\icmlkeywords{financial forecasting, large language models, reflexivity,
market regimes, boom-bust cycles, prompt engineering}
\vskip 0.3in
]
\printAffiliationsAndNotice{}

\begin{abstract}
We study how frontier large language models (LLMs) behave as
financial forecasters during boom-bust market cycles when made
progressively aware of Soros's theory of reflexivity.  Standard
AI-assisted forecasting treats the market as an exogenous system.
Reflexivity theory holds otherwise: prices shape fundamentals, and
every forecaster is a participative agent in the loop it analyzes.
We evaluate three frontier models --- GPT-5, Claude Sonnet~4.6, and
Gemini~3~Pro --- under four accumulating zero-shot conditions across
two historically distinct episodes: the dot-com bubble (1996--2001)
and the global financial crisis (2004--2009).  The primary metric is
directional forecasting accuracy; we also report the Sharpe ratio of
an implied long/cash strategy to capture the risk-adjusted economic
value of the forecasts.  All inputs are anonymized and normalized to
guard against memorization.  We find that conditions incorporating
reflexivity awareness improve forecasting accuracy differently across models and context windows, revealing that the same theoretical
awareness can produce qualitatively different forecasting behavior
across frontier LLMs.
\end{abstract}

\section{Introduction}
\label{sec:intro}

Standard approaches to AI-assisted financial forecasting treat the
market as an exogenous object --- a system whose past behavior
contains signals about its future states.  But financial markets have
a property that distinguishes them from most prediction targets: the
participants who form beliefs about the market are the same agents
whose actions move it.  A forecaster is not a neutral observer; it
is, in a meaningful sense, part of the phenomenon it is predicting.

George Soros's theory of \emph{reflexivity}~\citep{soros1987,soros1998}
formalizes this property.  He argues that the relationship between
prices and fundamentals is bidirectional: prices influence investor
perception, which drives capital allocation, which alters the
fundamentals that prices are supposed to reflect.  This
self-reinforcing loop generates a \emph{prevailing bias} --- a
systematic deviation of prices from what earnings alone would justify
--- that can grow over extended periods before violently reversing.

Soros further observed that the reflexive process tends to unfold
in a characteristic boom-bust sequence, tracked through the level
and momentum of the price-to-earnings (P/E) ratio~\citep{soros1995}.
The cycle moves from an \textit{Unrecognized} nascent uptrend through
\textit{InitialPhase} and \textit{Acceleration} in the boom arc,
before reversing through \textit{Twilight}, \textit{Crash}, and
\textit{Recovery} in the bust arc.  This stage vocabulary provides a
structured lens through which an LLM can reason about the current
state of a market cycle.

This paper asks a focused empirical question: \emph{how does
awareness of the reflexivity mechanism change LLM directional
forecasting during boom-bust cycles?}  We investigate this by
progressively revealing the theory to the model --- from price
patterns and earnings fundamentals alone, through the reflexivity
mechanism, to a full stage-by-stage account of how the
self-reinforcing loop operates --- and measuring the effect on
directional accuracy and risk-adjusted returns across two
historically distinct episodes.

\section{Background}
\label{sec:background}

\subsection{The Theory of Reflexivity and the Boom-Bust Cycle}

\citet{soros1987} argues that two functions operate simultaneously in
financial markets.  The \emph{cognitive function} runs from reality
to views: participants observe prices and fundamentals and form
beliefs, but those beliefs are always incomplete and biased.  The
\emph{participative function} runs in reverse: participants act on
their beliefs through buying, selling, and capital allocation, and
those actions change the very fundamentals they were trying to assess.
A rising price can genuinely improve a firm's access to capital,
strengthen its balance sheet, and attract investment --- making
fundamentals better than they would otherwise have been, which in
turn seems to validate the higher price.

This two-way causality produces a \emph{prevailing bias}: a
collectively reinforced misconception about where prices should be,
measurable as the gap between the actual P/E ratio and its long-run
mean.  When the bias and the underlying trend reinforce each other,
the system moves progressively further from equilibrium.  When the
accumulated deviation becomes unsustainable, the same feedback
mechanism reverses --- and because the bias must fully unwind, busts
are typically faster and steeper than the preceding
booms~\citep{soros1998}.

Soros further observed that the reflexive process tends to follow a
characteristic boom-bust sequence~\citep{soros1995}.  We
operationalize this sequence as seven stages, named and characterized
through the observable dynamics of price and EPS.  The \emph{boom
arc} builds from an Unrecognized trend through InitialPhase, Testing,
and Acceleration, with the P/E ratio rising progressively further
above its historical mean as the self-reinforcing loop intensifies.
The \emph{bust arc} mirrors this in reverse: Twilight marks the
stall in momentum while EPS remains intact; Crash arrives when
falling prices feed back into fundamentals and EPS begins to
deteriorate; and Recovery sees the bust run its course as the
prevailing bias collapses.  The stage names and their signal-based
criteria --- grounded in the price and EPS series --- are our
operationalization of Soros's framework, and are related to the
broader narrative economics literature~\citep{shiller2019,akerlofshiller2009}.

\subsection{LLMs in Financial Forecasting}

LLMs have demonstrated useful signal in financial domains.
\citet{lopezlira2023} show that LLM-generated sentiment predicts
stock returns; \citet{matera2025} finds that narratives from
earnings calls improve prediction of analyst expectations and
realized earnings. \citet{park2024} and \citet{yang2025} apply LLM
agents to anomaly detection and market simulation.  A key gap
identified by \citet{bondklokzhu2023} is that most LLM forecasting
pipelines do not condition on macroeconomic regime, yielding fragile
out-of-sample performance.  Our work differs in two ways: we work
from quantitative time series rather than text, and our central
intervention is theoretical awareness of market structure rather
than information augmentation.

\section{Experimental Framework}
\label{sec:framework}

\paragraph{Data.}
We use monthly adjusted closing prices for the S\&P~500 index from
Yahoo Finance and monthly trailing twelve-month earnings-per-share
(EPS) from Robert Shiller's publicly available dataset~\citep{shillerdata},
beginning January~1986 to provide a full 10-year history for the
rolling P/E z-score normalization underlying the stage definitions.

\paragraph{Evaluation episodes.}
We evaluate on two historically significant boom-bust cycles, each
spanning approximately 72 months:

\begin{itemize}\setlength\itemsep{2pt}
  \item \textbf{Dot-com bubble} (Jan~1996--Dec~2001): a speculative
  cycle driven by narrative excess and extreme P/E expansion in
  technology.  The prevailing bias in equities is the primary signal,
  making this a textbook reflexivity test case.

  \item \textbf{Global financial crisis} (Jan~2004--Dec~2009): a
  structural credit cycle where reflexive feedback operated through
  financial system fundamentals rather than equity P/E directly.
  This tests whether reflexivity-grounded reasoning generalizes
  beyond the canonical speculative bubble.
\end{itemize}

Evaluating on both episodes together yields an overall estimate;
evaluating separately reveals whether reflexivity awareness
generalizes across structurally different cycle types.

\paragraph{Prompting conditions.}
We evaluate four accumulating zero-shot conditions; full prompt text
is in Appendix~\ref{app:prompts}.  Soros's seven-stage boom-bust
framework is given to all conditions as instructional context.  What
varies is the theoretical framing of \emph{why} stages arise:
(A)~price only; (B)~adds EPS; (C)~adds the reflexivity mechanism
(cognitive and participative functions, prevailing bias); (D)~adds
how that mechanism operates at each specific stage.

\paragraph{Tasks and metrics.}
At each evaluation window, the LLM is asked to forecast three
quantities for the next monthly close: (i)~the boom-bust cycle
\emph{phase}, to prompt the model to reason explicitly about where
the market stands in the reflexive cycle; (ii)~the \emph{direction}
of the price return; and (iii)~the \emph{return magnitude} as a signed percentage, to
encourage calibrated probabilistic thinking.

Our evaluation focuses on the directional forecast.  The primary
metric is \textbf{directional accuracy}: the fraction of months on
which the predicted direction matches the realized direction.  We
also report the \textbf{annualized Sharpe ratio} of an implied
long/cash strategy (position $s_t \in \{1, 0.5, 0\}$ for
\texttt{up}/\texttt{neutral}/\texttt{down}; portfolio return
$r_t^{\mathrm{port}} = s_t \cdot r_{t+1}$), detailed in
Appendix~\ref{app:sharpe}.  Directional accuracy treats all months
equally, whereas Sharpe is return-weighted and therefore reflects
whether correct calls are concentrated in the economically most
significant months --- a distinction that matters greatly in markets
characterized by rare but extreme moves.

\paragraph{Models.}
We evaluate three frontier large language models: \textbf{GPT-5}
(temperature~$= 1$, the API default; adjustment is not permitted),
\textbf{Claude Sonnet~4.6} (temperature~$= 0$), and
\textbf{Gemini~3~Pro} (temperature~$= 0$).  All experiments use
zero-shot prompting.

\paragraph{Context window.}
We evaluate two context window lengths, $W \in \{36, 60\}$ months,
to assess whether the length of observable history moderates the
effect of theoretical scaffolding.  

\paragraph{Normalization and knowledge-cutoff treatment.}
Both evaluation episodes predate all three models' training cutoffs.
To guard against retrieval of memorized outcomes, all inputs are
normalized using a shared base.  Let $P_0$ denote the price at the
first month of the context window.  The normalized price series is
$P_t' = 100 \cdot P_t / P_0$ and the normalized EPS series is
$\mathrm{EPS}_t' = 100 \cdot \mathrm{EPS}_t / P_0$ --- both divided
by the \emph{same} base price, so that $P_t' / \mathrm{EPS}_t' = P_t
/ \mathrm{EPS}_t$ and the true P/E ratio is preserved exactly.
Dates are replaced by $t = 1, \ldots, T$ and the index is labeled
``equity market index,'' removing all absolute level and calendar
information while preserving the economically meaningful P/E signal.
\section{Results}
\label{sec:results}

\paragraph{Directional Accuracy and Sharpe Ratio} Table~\ref{tab:accuracy} reports directional accuracy across episodes,
models, and context windows ($n = 72$ per episode cell; $n = 144$
combined).  The Sharpe ratios corresponding to the same cells are in
Appendix~\ref{app:sharpe}.  Taken together, the results show that
reflexivity awareness can improve LLM forecasting, but the benefit
is heterogeneous across models, context windows, and episode type ---
a pattern that itself constitutes a substantive finding.

\paragraph{Gemini benefits from theory across both windows.}
Gemini~3~Pro is the most consistent beneficiary of theoretical
scaffolding.  At $W = 36$, Condition~B already lifts directional
accuracy to 62.5\% in the GFC (Cond.~A: 56.9\%), and Conditions C
and D maintain gains in both episodes; the combined Sharpe rises
monotonically from $-0.024$ (A) to $+0.277$ (D).  At $W = 60$, the
benefit is present in both episodes: GFC accuracy reaches 59.7\%
under C and D, and the combined Sharpe under C is $+0.357$.  Gemini
thus demonstrates that awareness of reflexive market dynamics --- both
the mechanism and the stage-level mapping --- can translate into more
accurate and better-calibrated directional forecasts.

\paragraph{GPT-5 benefits only at the longer context window.}
GPT-5 presents the starkest context-window dependence in the dataset.
At $W = 36$, directional accuracy declines monotonically from
Condition~A (54.9\% combined) to D (50.7\%), the only result
consistently near or below the 50\% random baseline.  At $W = 60$,
the pattern reverses completely: accuracy rises from 52.8\% (A) to
59.0\% (D) combined, and the Sharpe ratio reaches $+0.440$ (D) with
an exceptional $+0.674$ in the GFC under Condition~C --- the highest
single-cell Sharpe in the dataset.  This reversal suggests that the
reflexivity framework requires sufficient P/E accumulation history
to be actionable.  With only 36 months of context, GPT-5 cannot
observe the full build-up of the prevailing bias and the theory
induces premature bearish positions; 60 months provides enough
signal for the framework to serve as productive scaffolding.

\paragraph{Claude shows limited benefit from theory.}
Claude Sonnet~4.6 does not benefit consistently from reflexivity
awareness.  At $W = 36$, Condition~B raises dot-com accuracy to a
dataset-high 58.3\% but the Sharpe trajectory is non-monotone:
$+0.454$ at B, falling to $+0.128$ at C, and only partially
recovering to $+0.256$ at D.  At $W = 60$, adding EPS or theory
reduces both accuracy and Sharpe relative to the naive baseline.
The core difficulty is timing: the reflexivity mechanism (C)
correctly identifies overvaluation but provides no context for
\emph{when} the reversal begins, inducing over-bearish positions
during extended Mania phases.  The stage-level mapping (D) partially
mitigates this by conveying that Mania can persist with upward
momentum, but the correction is incomplete.

\paragraph{Episode type moderates the theory benefit.}
Across all models, reflexivity conditions produce more consistent
gains in the GFC than in the dot-com bubble.  In the GFC, the
reflexive feedback between falling asset prices and deteriorating
bank fundamentals directly matches the theoretical mechanism,
providing concrete traction.  In the dot-com, the Mania phase was
exceptionally prolonged and narrative-driven, making overvaluation
signals misleading for timing.  The episode dependence implies that
LLMs --- like human traders --- have context-specific strengths and
weaknesses: a model aware of reflexivity may be well-equipped for
credit-cycle busts but poorly calibrated for speculative manias.

\begin{table*}[t]
\centering
\caption{Directional forecasting accuracy. $W \in \{36, 60\}$ represents context window; $n = 72$ months per episode, $n = 144$ combined.
         Best per episode $\times$ model $\times$ context window in \textbf{bold}.
         }
\label{tab:accuracy}
\small
\begin{tabular}{ll ccc ccc}
\toprule
& &
  \multicolumn{3}{c}{\textbf{Directional accuracy ($W=60$)}} &
  \multicolumn{3}{c}{\textbf{Directional accuracy ($W=36$)}} \\
\cmidrule(lr){3-5}\cmidrule(lr){6-8}
\textbf{Episode} & \textbf{Condition}
  & \textbf{GPT-5} & \textbf{Claude 4.6} & \textbf{Gemini 3 Pro}
  & \textbf{GPT-5} & \textbf{Claude 4.6} & \textbf{Gemini 3 Pro} \\
\midrule
\multirow{4}{*}{\parbox{1.5cm}{Dot-com\\1996--2001}} & (A) Naive & 0.514 & \textbf{0.528} & 0.542 & 0.486 & 0.569 & 0.528 \\
  & (B) + EPS & 0.500 & 0.514 & 0.569 & \textbf{0.514} & \textbf{0.583} & 0.486 \\
  & (C) + Reflexivity & 0.528 & 0.486 & 0.542 & \textbf{0.514} & 0.528 & 0.528 \\
  & (D) Full Theory & \textbf{0.583} & 0.486 & \textbf{0.597} & 0.500 & 0.556 & \textbf{0.542} \\
\addlinespace
\multirow{4}{*}{\parbox{1.5cm}{GFC\\2004--2009}} & (A) Naive & 0.542 & \textbf{0.597} & 0.556 & \textbf{0.611} & 0.556 & 0.569 \\
  & (B) + EPS & 0.556 & 0.542 & 0.583 & 0.569 & 0.542 & \textbf{0.625} \\
  & (C) + Reflexivity & \textbf{0.597} & 0.569 & \textbf{0.597} & 0.556 & 0.556 & 0.611 \\
  & (D) Full Theory & \textbf{0.597} & 0.569 & 0.583 & 0.514 & \textbf{0.583} & 0.597 \\
\addlinespace
\multirow{4}{*}{\parbox{1.5cm}{Overall\\combined}} & (A) Naive & 0.528 & \textbf{0.562} & 0.549 & \textbf{0.549} & 0.562 & 0.549 \\
  & (B) + EPS & 0.528 & 0.528 & 0.576 & 0.542 & 0.562 & 0.556 \\
  & (C) + Reflexivity & 0.562 & 0.528 & 0.569 & 0.535 & 0.542 & \textbf{0.569} \\
  & (D) Full Theory & \textbf{0.590} & 0.528 & \textbf{0.590} & 0.507 & \textbf{0.569} & \textbf{0.569} \\
\bottomrule
\end{tabular}
\end{table*}

\section{Discussion}
\label{sec:discussion}

\paragraph{Reflexivity awareness benefits LLMs, but unevenly.}
The central finding is that making an LLM aware of the reflexivity
mechanism can improve directional forecasting, but the effect is
heterogeneous across models, context windows, and market episodes.
Gemini benefits broadly; GPT-5 benefits only when given a longer
historical window; Claude shows limited and inconsistent gains.
This heterogeneity is not noise --- it is itself informative.  It
suggests that each model arrives with an internalized prior about
financial markets, and that the same theoretical scaffold interacts
differently with those priors.  Reflexivity-aware prompting is not a
uniformly beneficial intervention; its value depends critically on
what the model already ``knows'' about market dynamics.

\paragraph{LLMs exhibit context-specific strengths, like human traders.}
The episode dependence of the theory benefit reveals something deeper
than a model calibration issue.  In the GFC, where reflexive feedback
directly linked falling asset prices to fundamental deterioration,
theory conditions consistently improve forecasting across models.  In
the dot-com bubble, where the Mania phase was prolonged and
narrative-driven, the same theory tends to induce premature bearish
positions that are costly in a sustained bull market.  This parallels
a well-known challenge for human traders: understanding that a market
is reflexively overvalued is not sufficient to time the reversal.
The finding suggests that LLMs, when equipped with economic theory,
exhibit analogous strengths and weaknesses --- better calibrated for
credit-cycle busts that follow the reflexive mechanism closely, and
less reliable in speculative manias where the prevailing bias can
persist far beyond what theory would predict.

\paragraph{Context window is a design variable, not a robustness check.}
The strong interaction between context window length and the effect of
theoretical scaffolding --- most dramatically for GPT-5, which
reverses from underperforming the random baseline at $W = 36$ to
achieving the best combined Sharpe at $W = 60$ --- implies that
context window length should be treated as a substantive design
choice.  The reflexivity framework describes a multi-year process of
bias accumulation and reversal; a 36-month window may provide
insufficient history to make the framework actionable, while 60
months allows the model to observe the full arc of P/E dynamics.

\paragraph{Limitations.}
The two episodes provide limited statistical power for cross-episode generalization. Future work should extend the evaluation to additional boom-bust
episodes to increase statistical power and test whether
the episode-type dependence documented here is robust.  A broader
sweep of context window lengths would also clarify the threshold at which reflexivity theory becomes actionable for different models.

\section{Conclusion}
\label{sec:conclusion}

We study how three frontier LLMs behave as financial forecasters
during boom-bust cycles when made progressively aware of Soros's
theory of reflexivity.  Our four-condition zero-shot ablation shows
that reflexivity awareness improves directional accuracy for Gemini
and GPT-5 (the latter strongly at $W = 60$ months) while producing
a non-monotone pattern for Claude --- where the reflexivity mechanism
alone (Condition C) reduces Sharpe by inducing premature bearish
positions in the Mania phase, and the full stage-level mapping
(Condition D) partially corrects this timing problem.

The accuracy--Sharpe divergence is a key methodological finding:
higher directional accuracy does not guarantee better risk-adjusted
returns, because accuracy treats all months equally while economic
performance is return-weighted.  A model that correctly calls the
direction in low-magnitude months while missing large Mania-phase
rallies can achieve high accuracy but negative Sharpe.

The substantial heterogeneity across models in response to the same
theoretical scaffold --- from strong benefits for Gemini to
context-dependent effects for GPT-5 to non-monotone Sharpe for Claude
--- suggests that reflexivity-aware prompting interacts with each
model's internalized priors about financial data.  Future work should
examine whether this heterogeneity tracks training data differences,
explore explicit self-referential framing (the forecaster as
participant), and extend the evaluation to additional market episodes.

\bibliography{reflexivity_llm}

@book{soros1987,
  author    = {Soros, George},
  title     = {The Alchemy of Finance},
  publisher = {Simon \& Schuster},
  year      = {1987}
}

@book{soros1995,
  author    = {Soros, George},
  title     = {Soros on Soros: Staying Ahead of the Curve},
  publisher = {Wiley},
  year      = {1995}
}

@book{soros1998,
  author    = {Soros, George},
  title     = {The Crisis of Global Capitalism: Open Society Endangered},
  publisher = {PublicAffairs},
  address   = {New York},
  year      = {1998}
}

@book{shiller2019,
  author    = {Shiller, Robert J.},
  title     = {Narrative Economics: How Stories Go Viral and Drive Major Economic Events},
  publisher = {Princeton University Press},
  year      = {2019}
}

@book{akerlofshiller2009,
  author    = {Akerlof, George A. and Shiller, Robert J.},
  title     = {Animal Spirits: How Human Psychology Drives the Economy and
               Why It Matters for Global Capitalism},
  publisher = {Princeton University Press},
  year      = {2009}
}

@article{lopezlira2023,
  author  = {Lopez-Lira, Alejandro and Tang, Yuehua},
  title   = {Can {ChatGPT} Forecast Stock Price Movements?
             {Return} Predictability and Large Language Models},
  journal = {SSRN Working Paper},
  year    = {2023},
  note    = {Available at SSRN: \url{https://ssrn.com/abstract=4412788}}
}

@article{matera2025,
  author  = {Matera, Giuseppe},
  title   = {Corporate Earnings Calls and Analyst Beliefs},
  journal = {arXiv preprint},
  volume  = {arXiv:2511.15214},
  year    = {2025}
}

@article{park2024,
  author  = {Park, Taejin},
  title   = {Enhancing Anomaly Detection in Financial Markets with an
             {LLM}-Based Multi-Agent Framework},
  journal = {arXiv preprint},
  volume  = {arXiv:2403.19735},
  year    = {2024}
}

@article{yang2025,
  author  = {Yang, Yuzhen and others},
  title   = {{TwinMarket}: A Scalable Behavioral and Social Simulation
             for Financial Markets},
  journal = {arXiv preprint},
  volume  = {arXiv:2502.01506},
  year    = {2025}
}

@article{bondklokzhu2023,
  author  = {Bond, Shaun A. and Klok, Hayden and Zhu, Min},
  title   = {Large Language Models and Financial Market Sentiment},
  journal = {SSRN Working Paper},
  year    = {2023},
  note    = {Available at SSRN: \url{https://ssrn.com/abstract=4584928}}
}

@misc{shillerdata,
  author       = {Shiller, Robert J.},
  title        = {Online Data: {U.S.} Stock Markets 1871--Present and {CAPE} Ratio},
  year         = {2025},
  howpublished = {\url{http://www.econ.yale.edu/~shiller/data.htm}},
  note         = {Accessed May 2025}
}
\bibliographystyle{icml2026}

\appendix

\section{Prompting Conditions: Block Composition}
\label{app:prompts}

Table~\ref{tab:prompt_structure} shows which blocks compose the system
prompt for each condition.  The full verbatim text of every block is
given in Appendix~\ref{app:prompt_blocks} at the end of this document.

\begin{table}[h]
\centering
\caption{System prompt block composition by condition.}
\label{tab:prompt_structure}
\scriptsize
\setlength{\tabcolsep}{4pt}
\begin{tabular}{p{3.4cm}cccc}
\toprule
\textbf{Block}              & \textbf{A} & \textbf{B} & \textbf{C} & \textbf{D} \\
\midrule
1 — Role + price series         & \checkmark & \checkmark & \checkmark & \checkmark \\
2 — EPS series                  & ---        & \checkmark & \checkmark & \checkmark \\
3 — Stage definitions           & \checkmark & \checkmark & \checkmark & \checkmark \\
4 — Reflexivity mechanism       & ---        & ---        & \checkmark & \checkmark \\
5 — Stage-level reflex.\ map    & ---        & ---        & ---        & \checkmark \\
\ \ Task + output schema        & \checkmark & \checkmark & \checkmark & \checkmark \\
\midrule
Approx.\ size & 1.5 kB & 1.6 kB & 4.3 kB & 10.0 kB \\
\bottomrule
\end{tabular}
\end{table}

\section{Sharpe Ratio Results}
\label{app:sharpe}

The annualized Sharpe ratio of the long/cash strategy is:
\begin{equation}
  \mathrm{SR} = \frac{\sqrt{12}\cdot
    \mathbb{E}[r^{\mathrm{port}}]}{\mathrm{std}[r^{\mathrm{port}}]},
  \label{eq:sr}
\end{equation}
where position $s_t \in \{1.0, 0.5, 0.0\}$ maps \texttt{up}/\texttt{neutral}/\texttt{down}
to portfolio return $r_t^{\mathrm{port}} = s_t \cdot r_{t+1}$.
The buy-and-hold benchmark sets $s_t \equiv 1.0$ always (SR~$= 0.35$
over the combined period).

Table~\ref{tab:sharpe} mirrors the structure of Table~\ref{tab:accuracy}
with Sharpe ratios in place of accuracy.  The starkest result is
GPT-5~C in the GFC at $W=60$: SR~$=+0.674$, far above the
buy-and-hold benchmark of 0.02.  In the dot-com episode, where the
buy-and-hold SR is 0.65, no condition approaches parity --- the
sustained Mania phase makes the market difficult to beat regardless
of theoretical awareness.  Claude achieves the highest dot-com Sharpe
(A, $W=36$: $+0.559$) by staying fully long in the bull market,
while Claude~D achieves the best GFC Sharpe ($+0.515$ at $W=36$).

\begin{table*}[t]
\centering
\caption{Annualized Sharpe ratio (Eq.~\ref{eq:sr}).
         $W \in \{36, 60\}$ represents context window; $n = 72$ months per episode, $n = 144$ combined.
         Best per episode $\times$ model $\times$ context window in \textbf{bold}. Buy-and-hold SR: Dot-com = $0.65$, GFC = $0.02$, Overall = $0.35$.}
\label{tab:sharpe}
\small
\begin{tabular}{ll ccc ccc}
\toprule
& &
  \multicolumn{3}{c}{\textbf{Sharpe ratio ($W=60$)}} &
  \multicolumn{3}{c}{\textbf{Sharpe ratio ($W=36$)}} \\
\cmidrule(lr){3-5}\cmidrule(lr){6-8}
\textbf{Episode} & \textbf{Condition}
  & \textbf{GPT-5} & \textbf{Claude 4.6} & \textbf{Gemini 3 Pro}
  & \textbf{GPT-5} & \textbf{Claude 4.6} & \textbf{Gemini 3 Pro} \\
\midrule
\multirow{4}{*}{\parbox{1.5cm}{Dot-com\\1996--2001}} & (A) Naive & 0.148 & \textbf{0.110} & 0.181 & \textbf{0.087} & \textbf{0.559} & 0.104 \\
  & (B) + EPS & -0.004 & 0.015 & 0.300 & 0.032 & 0.454 & -0.006 \\
  & (C) + Reflexivity & 0.135 & -0.066 & 0.230 & 0.080 & 0.128 & 0.161 \\
  & (D) Full Theory & \textbf{0.361} & 0.002 & \textbf{0.336} & 0.086 & 0.256 & \textbf{0.166} \\
\addlinespace
\multirow{4}{*}{\parbox{1.5cm}{GFC\\2004--2009}} & (A) Naive & 0.117 & 0.365 & -0.094 & \textbf{0.262} & 0.032 & -0.172 \\
  & (B) + EPS & 0.461 & 0.020 & 0.317 & 0.007 & 0.408 & 0.395 \\
  & (C) + Reflexivity & \textbf{0.674} & \textbf{0.471} & \textbf{0.503} & -0.096 & 0.324 & 0.261 \\
  & (D) Full Theory & 0.531 & 0.121 & 0.363 & 0.096 & \textbf{0.515} & \textbf{0.404} \\
\addlinespace
\multirow{4}{*}{\parbox{1.5cm}{Overall\\combined}} & (A) Naive & 0.134 & \textbf{0.229} & 0.052 & \textbf{0.169} & 0.311 & -0.024 \\
  & (B) + EPS & 0.211 & 0.017 & 0.308 & 0.020 & \textbf{0.433} & 0.181 \\
  & (C) + Reflexivity & 0.383 & 0.183 & \textbf{0.357} & -0.002 & 0.219 & 0.208 \\
  & (D) Full Theory & \textbf{0.440} & 0.058 & 0.349 & 0.091 & 0.377 & \textbf{0.277} \\
\bottomrule
\end{tabular}
\end{table*}




\section{Full System Prompt Blocks}
\label{app:prompt_blocks}

All text is delivered verbatim as the \texttt{system} message; the
user message contains only the normalized time series.  Block
composition by condition is in Table~\ref{tab:prompt_structure}
(Appendix~\ref{app:prompts}).

\clearpage
\onecolumn

\begin{tcolorbox}[
  width=0.95\textwidth, colback=gray!10, colframe=darkgray,
  title=Block 1 — Role and price series \ (all conditions),
  coltitle=white, boxsep=2pt, left=5pt, right=5pt, top=5pt, bottom=5pt,
  fonttitle=\small\bfseries, fontupper=\footnotesize\ttfamily
]
You are an expert financial analyst examining a normalized price index for
an equity market.

You will receive a monthly time series indexed t=1 (oldest) to t=T (most
recent), where the price is normalized to 100 at t=1.
\end{tcolorbox}

\smallskip

\begin{tcolorbox}[
  width=0.95\textwidth, colback=gray!10, colframe=darkgray,
  title=Block 2 — EPS series \ (Conditions B\, C\, D),
  coltitle=white, boxsep=2pt, left=5pt, right=5pt, top=5pt, bottom=5pt,
  fonttitle=\small\bfseries, fontupper=\footnotesize\ttfamily
]
You also receive an earnings-per-share (EPS) index for the same market.
Both the price index and the EPS index are divided by the same base value
(the price at t=1), so their ratio at any point equals the true P/E ratio:

\medskip
\textbf{P/E at t = PRICE INDEX[t] / EPS INDEX[t]}
\end{tcolorbox}

\smallskip

\begin{tcolorbox}[
  width=0.95\textwidth, colback=gray!10, colframe=darkgray,
  title=Block 3 — Seven-stage boom-bust cycle definitions \ (all conditions),
  coltitle=white, boxsep=2pt, left=5pt, right=5pt, top=5pt, bottom=5pt,
  fonttitle=\small\bfseries, fontupper=\footnotesize\ttfamily,
  breakable
]
\textbf{Stage 1 — Unrecognized.}
A genuine improvement in fundamentals begins. EPS is growing. Price starts
to rise and outpace EPS slightly, but the divergence is small and goes
unrecognized by most participants. The P/E ratio is near or just below its
long-run mean.

\medskip\textbf{Stage 2 — InitialPhase.}
The trend becomes recognized and the recognition reinforces it. Price-EPS
divergence begins to widen. The P/E ratio rises above its long-run mean.
Not yet far-from-equilibrium --- the self-reinforcing loop is engaging but
has not accelerated.

\medskip\textbf{Stage 3 — Testing.}
A price correction pulls back at least 5\% from a recent peak. EPS
continues to grow --- this is a price-only correction. If price recovers
to a new high, the trend emerges strengthened. Identified retrospectively;
two test periods may occur.

\medskip\textbf{Stage 4 — Acceleration.}
The P/E ratio is clearly elevated above its long-run mean and rising. EPS
is still growing but price has moved far ahead of fundamentals. The
self-reinforcing loop is at its most powerful. Ends at the price peak ---
the moment of truth --- after which price can no longer be sustained.

\medskip\textbf{Stage 5 — Twilight.}
Price has reversed from its peak and is falling or flattening.
\textbf{Critical signal: EPS has NOT yet started declining.}
The trend may be sustained by inertia but ceases to be reinforced by belief.

\medskip\textbf{Stage 6 — Crash.}
The falling price feeds back into fundamentals ---
\textbf{EPS begins to decline.}
Both price and EPS are now falling in a self-reinforcing bust loop.
The cycle's characteristic rapid collapse occurs here.

\medskip\textbf{Stage 7 — Recovery.}
The bust has run its course. Price begins recovering but EPS may still be
declining. Mirrors Stage 1: a new uptrend emerges but goes unrecognized
by most participants. P/E near or below long-run mean.

\medskip\textit{Important: the process may be aborted at any stage.
The model describes the complete case; in practice, external shocks
may interrupt the sequence.}
\end{tcolorbox}

\smallskip

\begin{tcolorbox}[
  width=0.95\textwidth, colback=gray!10, colframe=darkgray,
  title=Block 4 — Theory of reflexivity \ (Conditions C\, D),
  coltitle=white, boxsep=2pt, left=5pt, right=5pt, top=5pt, bottom=5pt,
  fonttitle=\small\bfseries, fontupper=\footnotesize\ttfamily,
  breakable
]
Reflexivity describes a two-way feedback loop between participants' thinking
and the situation they are in. In financial markets, participants' biased
views drive their actions, and those actions change the very fundamentals
they are trying to assess.

\medskip Two functions operate simultaneously:

\medskip\textbf{Cognitive function (reality $\to$ views):}
Participants form beliefs about market prices, but the reality they observe
has already been shaped by prior beliefs and actions. Their view is always
incomplete and biased.

\medskip\textbf{Participative function (views $\to$ reality):}
Participants act on their biased views. Buying and selling moves prices.
Price movements alter the fundamentals --- a rising price improves a firm's
access to capital and strengthens its finances; a falling price does the
reverse.

\medskip The \textbf{prevailing bias} is the net gap between prices and
what the fundamentals alone would justify, observable as the gap between
the actual P/E ratio and its long-run historical mean.

\medskip Keep this theory in mind when examining the data. Use it to
interpret what the series signals about the interaction between
participants' beliefs and the fundamentals at each point in time.
\end{tcolorbox}

\smallskip

\begin{tcolorbox}[
  width=0.95\textwidth, colback=gray!10, colframe=darkgray,
  title=Block 5 — Stage-level reflexivity map \ (Condition D only),
  coltitle=white, boxsep=2pt, left=5pt, right=5pt, top=5pt, bottom=5pt,
  fonttitle=\small\bfseries, fontupper=\footnotesize\ttfamily,
  breakable
]
\textbf{Stage 1 — Unrecognized.}
Cognitive function lags reality: participants attribute the price rise to
noise. Participative function is weak. The prevailing bias is nascent; P/E
barely moves.

\medskip\textbf{Stage 2 — InitialPhase.}
Cognitive function catches up: the trend is recognized. Recognition
triggers the participative function --- new buyers push prices higher,
improving corporate access to capital, genuinely improving EPS, and
validating the higher P/E. The feedback loop is fully engaged.

\medskip\textbf{Stage 3 — Testing.}
A shock temporarily disrupts the loop. Most participants' cognitive
function still holds the bias intact; EPS continues to grow. The
participative function re-engages as buyers return on the dip. If price
recovers, the loop emerges strengthened.

\medskip\textbf{Stage 4 — Acceleration.}
Cognitive function pushes the bias to an extreme --- the trend feels
unshakable. Participative function is at maximum power: higher prices
sustain EPS growth and justify ever-higher P/E ratios. The boom builds
toward the price peak.

\medskip\textbf{Stage 5 — Twilight.}
Price peak is past. Cognitive function begins to crack --- bias ceases to
be reinforced by belief. Participative function weakening, but
\textbf{EPS has NOT yet been damaged} --- fundamentals still intact even
as price falls.

\medskip\textbf{Stage 6 — Crash.}
Cognitive function inverts --- extreme pessimism replaces the bias.
Participative function operates in reverse: falling prices destroy credit
availability, curtail investment, and directly reduce EPS (reflexive
fundamental damage). Catastrophic acceleration ensues.

\medskip\textbf{Stage 7 — Recovery.}
Prevailing pessimism begins to dissolve. Participative function's
destructive impact winds down: depressed prices attract value investors,
stabilizing fundamentals at the margin. As in Stage 1, most participants
do not yet recognize the new uptrend.
\end{tcolorbox}

\smallskip

\begin{tcolorbox}[
  width=0.95\textwidth, colback=gray!10, colframe=darkgray,
  title=Output schema \ (all conditions),
  coltitle=white, boxsep=2pt, left=5pt, right=5pt, top=5pt, bottom=5pt,
  fonttitle=\small\bfseries, fontupper=\footnotesize\ttfamily
]
\{

\quad "stage":\ \ \ \ "<Unrecognized | InitialPhase | Testing |

\qquad\qquad\qquad\quad Acceleration | Twilight | Crash | Recovery>",

\quad "direction": "<up | neutral | down>",

\quad "magnitude": "<signed monthly return in \%, e.g. +2.1 or -0.8>",

\quad "rationale": "<2--3 sentences>"

\}
\end{tcolorbox}

\clearpage
\twocolumn
\end{document}